\documentclass[aps,prd,11pt,a4paper,twocolumn,groupedaddress,preprintnumbers,nofootinbib,floatfix]{revtex4-1}

\usepackage[top=2.9cm, bottom=2.1cm, left=2cm, right=2cm]{geometry}
\usepackage{comment}
\usepackage{graphicx}
\usepackage{url}
\usepackage[bookmarks, pagebackref=false]{hyperref}
\usepackage[usenames,dvipsnames]{xcolor}
\usepackage{dcolumn}
\usepackage{bm}
\usepackage{bbm}
\usepackage{amsmath,amssymb,amsfonts}
\usepackage{color}
\usepackage{hyperref}
\usepackage[Symbolsmallscale]{upgreek}
\usepackage{amsmath}
\usepackage{amsfonts}
\usepackage{amssymb,dsfont}
\usepackage[vcentermath]{youngtab}
\usepackage[all]{xy}
\usepackage{pstricks}
\usepackage{dsfont}%
\usepackage{cases}
\usepackage{placeins}
\usepackage{xspace}
\usepackage{cancel} 
\usepackage{slashed}
\usepackage[caption=false, labelformat=simple, listofformat=subsimple, labelfont=default, margin=5pt,
    justification=raggedright]{subfig}

\usepackage{cases}
\usepackage{comment}
\usepackage{fixme}

\usepackage{placeins}
\usepackage{xspace}
\usepackage{cancel} 

\usepackage{siunitx}	
\usepackage{tikz}
\usetikzlibrary{shapes.callouts}

\newcommand{\beq}{\begin{eqnarray}}
\newcommand{\eeq}{\end{eqnarray}}

\newcommand{\bmp}{\noindent\begin{minipage}{16cm}}
\newcommand{\emp}{\end{minipage}\vskip 7mm} 

\newcommand{\eff}{\textrm{eff}}

    \newcommand{\ii}{\mathrm{i}}
    \newcommand{\dd}{\mathrm{d}}
    \newcommand{\uu}{\mathrm{u}}
    \newcommand{\ee}{\mathrm{e}}
    \newcommand{\Pf}{\mathrm{Pf}}
    \newcommand{\EW}{\mathrm{EW}}
    
    \newcommand{\SU}{\mathrm{SU}} 
    \newcommand{\Sp}{\mathrm{Sp}}
    \newcommand{\Tr}{\mathrm{Tr}}
    \newcommand{\LL}{\mathrm{L}}
    \newcommand{\RR}{\mathrm{R}}
    \newcommand{\SF}[1]{{\bf #1}}
    
    \newcommand{\Ssing}{\SF{\Omega}}
    \newcommand{\sing}{\omega}

    \newcommand{\abs}[1]{\left| #1 \right|}
    \newcommand{\brackets}[1]{\left\langle #1 \right\rangle}
    \newcommand{\hc}{\, + \, \text{h.c.} \,}

\def\lsim{\mathrel{\rlap{\lower4pt\hbox{\hskip1pt$\sim$}}
    \raise1pt\hbox{$<$}}}                
\def\gsim{\mathrel{\rlap{\lower4pt\hbox{\hskip1pt$\sim$}}
    \raise1pt\hbox{$>$}}}                

\begin{document}

\title{Raising the SUSY-breaking scale in a Goldstone-Higgs model}
\author{Tommi {\sc Alanne}}
\email{alanne@cp3.sdu.dk}
\affiliation{{CP}$^{ \bf 3}${-Origins} \& the Danish Institute for Advanced Study {\rm{Danish IAS}},  University of Southern Denmark, Campusvej 55, DK-5230 Odense M, Denmark.}
\author{Heidi {\sc Rzehak}}
\email{rzehak@cp3.sdu.dk}
\affiliation{{CP}$^{ \bf 3}${-Origins} \& the Danish Institute for Advanced Study {\rm{Danish IAS}},  University of Southern Denmark, Campusvej 55, DK-5230 Odense M, Denmark.}
\author{Francesco {\sc Sannino}}
\email{sannino@cp3.dias.sdu.dk}
\affiliation{{CP}$^{ \bf 3}${-Origins} \& the Danish Institute for Advanced Study {\rm{Danish IAS}},  University of Southern Denmark, Campusvej 55, DK-5230 Odense M, Denmark.}
\author{Anders Eller {\sc Thomsen}}
\email{aethomsen@cp3.sdu.dk}
\affiliation{{CP}$^{ \bf 3}${-Origins} \& the Danish Institute for Advanced Study {\rm{Danish IAS}},  University of Southern Denmark, Campusvej 55, DK-5230 Odense M, Denmark.}

\begin{abstract}
\noindent
We show that by combining the elementary-Goldstone-Higgs scenario and supersymmetry it is possible to raise the scale of supersymmetry breaking to several TeVs by relating it to the spontaneous-symmetry-breaking one.  This is achieved by first enhancing the global symmetries of the super-Higgs sector to $\SU(4)$ and then embedding the electroweak sector and the Standard-Model fermions.  We  determine the conditions under which the model achieves a vacuum such that the resulting Higgs is a pseudo-Goldstone boson. The main results are: the supersymmetry-breaking scale is identified with the spontaneous-symmetry-breaking scale of $\SU(4)$ which is several TeVs above the radiatively induced electroweak scale; intriguingly the global symmetry of the Higgs sector predicts the existence of two super-Higgs multiplets with one  mass eigenstate 
playing the role of the pseudo-Goldstone Higgs;  
the symmetry-breaking dynamics fixes 
$\tan \beta = 1$ and requires a supplementary singlet chiral superfield. 
We finally discuss the spectrum of the model that now features superpartners of the Standard-Model fermions and gauge bosons in the multi-TeV range.  \\
\\
[.1cm]
{\footnotesize  \it Preprint: CP$^3$-Origins-2016-026 DNRF90}
\end{abstract}
\maketitle 
\newpage
 

\section{Motivation}


\noindent
The two time-honored proposals to solve the naturalness problem of the Standard Model (SM) are either to invoke supersymmetry (SUSY) or to introduce new composite dynamics. 
The minimal supersymmetric extension of the SM (MSSM), however, has to confront the lack of experimental evidence for (light) superpartners of the SM fields~\cite{Aad:2015baa,CMS:2015ebf}.
This either pushes the SUSY-breaking scale far from the Fermi scale, or requires specific hierarchies among the squark generations, such as the presence of a light top squark and at least the first generation of squarks being rather heavy~\cite{Belanger:2015vwa}.

Our approach is to explain the large separation of Fermi and SUSY-breaking scales by constructing a template, where the Fermi scale is radiatively generated and the light observed Higgs boson is a pseudo-Goldstone boson (pGB) gaining all its mass radiatively. In this scenario, the electroweak symmetry breaking (EWSB) and the SUSY breaking can have origin at the same scale. In order to achieve the correct vacuum structure, we need to introduce a gauge-singlet chiral field, and the particle content of the model ends up being fairly similar to the next-to-minimal supersymmetric Standard Model (NMSSM), though with two additional chiral fields~\cite{Ellwanger:2009dp}. This is mostly coincidental as the reasoning for including the chiral field is very different. In spite of these similarities, the model stands out in that it predicts that the two Higgs doublets must be very close to the decoupling limit.   

The idea of raising the fundamental scale above the radiatively induced Fermi scale with the observed Higgs boson as an elementary pGB was put forward in 
non-supersymmetric framework in~\cite{Alanne:2014kea,Gertov:2015xma}, and it was further studied in relation to unification scenarios in~\cite{Alanne:2015fqh}.  In this work we upgrade the elementary Goldstone Higgs framework to a supersymmetric one. Models involving pGB-Higgs scenarios have been investigated earlier in the literature also in the context of SUSY ~\cite{Bae:2012ir,Birkedal:2004xi,Chankowski:2004mq,Berezhiani:2005pb,Roy:2005hg,Csaki:2005fc,Falkowski:2006qq,Bellazzini:2008zy,Bellazzini:2009ix,Kaminska:2010pc}. The main difference with respect to earlier investigations is that we determine the vacuum alignment of the model at the quantum level and thereby showing that the SUSY-breaking 
scale can be raised to the multi-TeV range. 

Concretely, we enlarge the global symmetry of the superpotential to $G$, and embed the electroweak (EW) symmetry as subgroup of $G$. As the global symmetry breaks spontanously at scale $v$, the EW symmetry breaks consequently depending on the alignment of the EW subgroup with respect
to the stability group of $G$. The alignment is in turn determined dynamically by quantum effects, which we will compute. In particular, if we denote the alignment
angle by $\phi$, the spontaneous breaking of $G$ generates the Fermi scale  $ v_{\EW} =v \sin \phi $ dynamically. If the dynamics of the model favour $ 0<\phi \ll 1 $, a large hierarchy between the scale of breaking $v$ and the Fermi scale is generated radiatively. Intriguingly soft SUSY-breaking operators are responsible for spontaneously breaking $ G $, and therefore $v$ is of the SUSY-breaking scale. This implies that the Fermi scale actually originates from spontaneous symmetry breaking near the (multi-) TeV SUSY-breaking scale. 

In the following, we present the minimal scenario  that involves the global symmetry breaking pattern 
$\SU(4)\rightarrow\Sp(4)$, and demonstrate that it is indeed possible to achieve $\phi\ll 1$. We present the ingredients of the model in 
Sec.~\ref{sec:SEGH}, and study the quantum corrections and vacuum alignment as well as the spectrum in Sec.~\ref{sec:quantumCorrections}.  Finally, we conclude in Sec.~\ref{sec:conc}.


\section{The model and notation}
\label{sec:SEGH}

	\noindent
	Following Ref. \cite{Alanne:2014kea}, we consider the spontaneous symmetry breaking pattern $\SU(4)\rightarrow\Sp(4)$. To this end we assume the Higgs sector to consist of a chiral superfield matrix, $ \SF{M} $, transforming according to the six-dimensional representation of $\SU(4)$. 	
	In terms of its constituent chiral superfields, the matrix $ \SF{M} $ reads
	\begin{equation}
		\SF{M} = \left[\tfrac{1}{\sqrt{2}} \SF{\Sigma} +2 \SF{\Pi}_iX^{i}\right]E,
		\label{eq:M}
	\end{equation}
	where $X^i$ with $i = 1, \dots, 5$ represent the broken generators associated with the quotient space of $\SU(4)\rightarrow\Sp(4)$, and the vacuum alignment is parameterized by the specific value of the antisymmetric matrix $E$. The explicit realization of the $\SU(4)$ generators was determined in \cite{Appelquist:1999dq}.  
	
	The components of the chiral superfields are
	\begin{align}
		\SF{\Sigma} &= \sigma(y) +\sqrt{2}\theta\psi_{\sigma}(y) + \theta^2 f_{\sigma}(y), \label{eq:Sigma_components}\\
		\SF{\Pi}_i &= \pi_i(y) + \sqrt{2}\theta\psi_i(y) + \theta^2 f_{i}(y), \label{eq:Pi_components}
	\end{align} 
	where $ y^{\mu} = x^{\mu} + i\theta \sigma^{\mu} \overline{\theta},  $ and $ \theta_{\alpha}, \overline{\theta}_{\dot{\alpha}} $
	are the Grassmannian superspace coordinates. The desired spontaneous symmetry breaking happens when the $ \sigma $ field acquires 
	a vacuum expectation value (vev).  
	
	The $\SU(4)$-symmetric superpotential for $ \SF{M} $ is uniquely determined by renormalizability and holomorphicity of the superpotential allowing for just one term: 
	\begin{equation}
		W_M = \mu\, \Pf(\SF{M}) = \frac{\mu}{2}\left(\SF{\Pi}_i^2 - \SF{\Sigma}^2 \right),
		\label{eq:WwithoutGamma}
	\end{equation}
	where $ \Pf $ denotes the Pfaffian of a matrix. This superpotential is essentially an $ \SU(4)$-symmetric version of the MSSM $ \mu $ term. Later we will see that this is insufficient for the model to produce the desired vacuum structure. Integrating out the auxiliary $ f $ fields results in the following contribution to the scalar potential: 
	\begin{equation}
		V_f = \abs{\mu}^2\left(\abs{\sigma}^2 + \abs{\pi_i}^2\right). \label{eq:Vf}
	\end{equation}

    \subsection{Electroweak and Yukawa sectors}
	
	\noindent
	The electroweak gauge group can be embedded in $\SU(4)$ in different ways with respect to the vacuum~\cite{Katz:2005au,Gripaios:2009pe,Galloway:2010bp,Barnard:2013zea,Ferretti:2013kya,Cacciapaglia:2014uja}.
	We parameterize this freedom  by 
	an angle $\phi$. The matrix $E$ in Eq.~\eqref{eq:M} is correspondingly replaced by $E_{\phi}$,
	\begin{equation}
	    E_{\phi}= \sin\phi 
	    \begin{pmatrix}
		0 & 1\\
		-1& 0\\
	    \end{pmatrix}
	    +\cos\phi 
	    \begin{pmatrix} 
		\ii \, \sigma_2 & 0\\
		0 & -\ii \, \sigma_2 \\
	    \end{pmatrix}.
	\end{equation}
	For $\phi=0$, the EW symmetry remains unbroken, while for $\phi=\pi/2$ it breaks directly to $\mathrm{U}(1)_Q$. The specific value of $\phi$ must be determined dynamically once the EW and top quantum corrections are taken into account. The Fermi scale, identified in the usual way from the gauge boson masses, is then $ v_{\mathrm{EW}} = v \sin \phi$ implying that for small $\phi$, the actual symmetry-breaking scale is significantly higher than the	EW scale. 
	
	Interestingly $ \SF{M} $ contains exactly two EW doublets, which without any further ado are identified with the two Higgs doublets required in the (minimal) supersymmetric Higgs mechanism:
		\begin{equation}
		\begin{split}
		&\SF{H}_{\uu}^{+} = \dfrac{\SF{\Pi}_1 + i\SF{\Pi}_2}{\sqrt{2}}, \quad \SF{H}_{\uu}^{0}=\dfrac{-\SF{K} + \SF{\Pi}_3}{\sqrt{2}},\\ 
		&\SF{H}_{\dd}^{0}=\dfrac{-\SF{K} - \SF{\Pi}_3}{\sqrt{2}}, \quad \SF{H}_{\dd}^{-} = \dfrac{\SF{\Pi}_1 - i\SF{\Pi}_2}{\sqrt{2}},
		\end{split}
		\label{eq:Higgses}
		\end{equation}
	with hypercharges $ +\tfrac{1}{2} $ and $ -\tfrac{1}{2}$, respectively, and $ \SF{K} = \sin \phi \SF{\Sigma} - i \cos \phi \SF{\Pi}_4 $. The realization of two different Higgs doublets in this model is thus not just motivated out of the necessity of being able to give masses to both up and down type fermions. Rather the requirement of having the EW symmetry embedded into the $ \SU(4) $ symmetry naturally gives two distinct doublets. 
	
	The electroweak gauge fields are implemented through four vector superfields, $\SF{W}^a$ and $\SF{B}$, as usual. We gauge the electroweak subgroup of $\SU(4)$ by introducing the gauge field 
		\begin{equation}
		\SF{G} = 2g\SF{W}^{a}T^{a}_{\LL} + 4g'Y\SF{B}T^{3}_{\RR},
		\end{equation}
	where $ Y=\tfrac{1}{2} $ is the hypercharge, and $T^a_{\LL},T^3_{\RR}$ are generators of the $\SU(2)_{\LL}$ and $\SU(2)_{\RR}$
	subgroups of $\SU(4)$, resp. The resulting gauge-invariant kinetic term for the $M$ sector is then
		\begin{equation}
		\mathcal{L}_{K} = \int \dd^2 \theta \; \dd^2 \overline{\theta}\; \tfrac{1}{2} \Tr\left[ \SF{M}^{\dagger} \ee^{\SF{G}} 
		\SF{M} \left(\ee^{\SF{G}}\right)^{\mathrm{T}} \right]. 
		\label{eq:L_kinetic}
		\end{equation}
	The gauge interactions will contribute to the scalar potential when integrating out the auxiliary $ d $ fields, giving   
	    \begin{align}
		V_d = &\tfrac{1}{8}\left(g^2+g'^2\right) \left(\abs{h_{\uu}^{+}}^2 + \abs{h_{\uu}^{0}}^2- \abs{h_{\dd}^{0}}^2
		- \abs{h_{\dd}^{-}}^2\right)^2 \nonumber\\ 
		&+ \tfrac{1}{2}g^2 \abs{h_{\uu}^{+\ast}h_{\dd}^{0} + h_{\uu}^{0\ast}h_{\dd}^{-}}^2, \label{eq:Vd2}
	    \end{align}
	where the $ h $ fields are the scalar components of the corresponding Higgs doublets. 
			
	With the identification of the Higgs doublets, we can include the Yukawa interactions by adding the following terms to the superpotential:
		\begin{equation}
		\begin{split}
		\label{eq:yukawaTerms}
		W_{\mathrm{Y}} =\  &y_{\mathrm{u}}^{ij} \SF{H}_{\uu}^{\alpha}\epsilon_{\alpha \beta}\SF{Q}_{i}^{\beta}\SF{U}_{j}
		- y_{\mathrm{e}}^{ij} \SF{H}_{\dd}^{\alpha}\epsilon_{\alpha \beta}\SF{L}_{i}^{\beta}\SF{E}_{j}\\
		&- y_{\mathrm{d}}^{ij} 
		\SF{H}_{\dd}^{\alpha}\epsilon_{\alpha \beta}\SF{Q}_{i}^{\beta}\SF{D}_{j},
		\end{split}
		\end{equation}
	where $\SF{Q}$, $\SF{L}$, $\SF{U}$, $\SF{D}$, $\SF{E}$ are the quark and lepton doublet and the up-type, down-type and lepton singlet superfields, $\alpha,\beta$ are $\SU(2)_{\mathrm{L}}$ indices, $ i,j $ refer to generations, and colour indices are left implicit. Both the $ d $-term contribution to the scalar potential and the Yukawa interactions are immediately recognizable, as they are identical to the MSSM counterparts \cite{Martin:1997ns}. This is not particularly surprising as the difference between this model and the MSSM lies in an extended Higgs-sector symmetry rather than in the gauge and fermion sectors.   
		
	Having identified the Higgs doublets in terms of their constituent fields in Eq.~\eqref{eq:Higgses}, it becomes clear that the sought-after vev in the $ \sigma $ direction inevitably yields $ \brackets{h_u^0} = \brackets{h_d^0} $. This corresponds exactly to $ \tan\beta=1 $ in terms of the usual ratio between the vevs of the two Higgs doublets implying the need for additional new physics at higher scales to avoid Landau poles. Nevertheless, the value of $\tan\beta$ is not a free parameter but instead a prediction of the model.	
    
    \subsection{Dynamics requires an additional singlet chiral superfield, $ \Ssing $} 

	\noindent
	So far we have considered a straightforward supersymmetric extension of the model presented in 
	\cite{Alanne:2014kea}. 
	The superpotential only contributes quadratic terms to the scalar potential of the $M$ sector, and thus the $ d $ terms are responsible for all possible quartic contributions. However, it is straightforward to show that the $ d $-term contribution to the scalar potential, Eq.~\eqref{eq:Vd2}, contains	no $ \abs{\sigma}^4 $ terms. Hence at the tree level, this minimal model is unable to give the desired spontaneous symmetry breaking pattern, $\SU(4)\rightarrow\Sp(4)$. Furthermore, soft SUSY-breaking terms are at most cubic in the scalar fields. The inclusion of such terms is therefore not sufficient to produce the desired vacuum structure. 

	A solution can, however, be found if we add a chiral superfield $ \Ssing = \sing(y) +\sqrt{2}\theta\psi_{\sing}(y) + \theta^2 f_{\sing}(y) $, which is an EW singlet. 
	This allows for a richer superpotential compared to that in 
	Eq.~\eqref{eq:WwithoutGamma}, and most significantly it provides new quartic terms to the scalar potential. 
	The most general $ \SU(4) $-symmetric superpotential terms containing the $\Ssing$ field are given by
	\begin{equation}
	    W_{\Ssing} = 2\lambda\, \Ssing \,\Pf(\SF{M}) +\tfrac{a}{3}\Ssing^{3} + \tfrac{b}{2}\Ssing^{2}
	    +c\, \Ssing.
	    \label{eq:full_W} 
	\end{equation}  
	With the inclusion of these terms to the superpotential, the $ f $-term contribution of Eq.~\eqref{eq:Vf} becomes
	\begin{equation}
	    \begin{split}
	    V_{f} = &\abs{a\sing^2 +b\sing + c + \lambda\left(\pi_i^2 -\sigma^2\right)}^{2} \\
		&+ \abs{2\lambda\sing + \mu}^2 \left(\abs{\sigma}^2 + \abs{\pi_i}^2\right).
	    \label{eq:Vf2}
	    \end{split}
	\end{equation}
	
	At first the inclusion of the gauge singlet $ \Ssing $ resembles the NMSSM \cite{Maniatis:2009re}. There is, however, an important difference in the motivation, as in the NMSSM the gauge singlet is included in order to avoid the $ \mu $ problem of the usual MSSM. In this model on the other hand, the $ \Ssing$ field is crucial to obtain the sought-after vacuum structure; without it, or some new mechanism altogether, it is not possible to achieve a vev in the desired $ \sigma $ direction.
	
	The general superpotential given in Eqs.~\eqref{eq:WwithoutGamma} and~\eqref{eq:full_W} has three dimensionful parameters: $ \mu $, $ b $, and $ c $.  However, these are not essential in obtaining the proclaimed properties of the model. Indeed for our numerical analysis of the model, the results of which are presented in section \ref{sec:quantumCorrections}, we have focused on the case where both $ \mu = b =0$, and $ c=0 $. This has the benefit of keeping the analysis minimal and can be justified by imposing a $ Z_3 $ symmetry. Furthermore it eliminates the introduction of new hierarchies among the different unprotected scales. 
    
    \subsection{Soft supersymmetry breaking}
	
	\noindent
	The tree-level scalar potential is given by the sum of the $ d $- and $ f $-term contributions given in Eqs. \eqref{eq:Vd2}, \eqref{eq:Vf2}\footnote{Additional $ f $-term contributions arise from the inclusion of the Yukawa interactions given in Eq. \eqref{eq:yukawaTerms}, though in the analysis we assume that the vev of the squark fields are vanishing.}.
	The minimum of this potential lies in the $\sing$ direction implying $ \brackets{h_\uu^{0}} = \brackets{h_\dd^{0}} = 0 $ and therefore no EWSB. Furthermore, the potential at the minimum vanishes leaving SUSY unbroken. 
	
	However given that supersymmetry must be broken at low energies, we now show that by adding soft SUSY-breaking terms we can achieve both, the desired pattern of chiral symmetry breaking and viable phenomenology.  

	For the sake of minimality, we consider soft-breaking terms that preserve the global $ \SU(4) $ symmetry 
	and do not contain the $\sing$ field. There are only two possible such terms given by 
	\begin{equation}
		V^M_{\mathrm{soft}} = \tfrac{1}{2}m^2 \Tr\left[M_s^{\dagger} M_s\right]+ \left(2\beta_{\mathrm{s}} \Pf(M_{\mathrm{s}}) \hc\right),
		\label{eq:softBreaking}
	\end{equation}
	where $ M_{\mathrm{s}} = \left[\tfrac{1}{\sqrt{2}} \sigma + 2 \pi_iX^{i}\right] E_{\phi} $ is the antisymmetric matrix consisting of the scalar fields of $ \SF{M} $.	Interestingly, in contrast to the MSSM, the soft masses of the two Higgs doublets do not have to differ in order to get a non-vanishing vev.
	
	Additional  soft-breaking terms must be included in the model to account for the mass splitting between the traditional matter and gauge fields and their respective supersymmetric partners. We have taken these to resemble the usual MSSM soft-breaking terms \cite{Martin:1997ns}. Although not a subject of investigation in this paper, this will naturally give a gluino mass in the multi-TeV range.   
	
	With the inclusion of the soft-breaking terms, the tree-level vacuum will in general give a non-zero vev for both $ \sing $ and $ \sigma $ fields. Note that as $ \sing $ is a gauge singlet, its vev will not influence the EWSB. The minimum of the tree-level potential is independent of the embedding angle $ \phi $. As a consequence, there is no preferred value of $ \phi $ at the tree level, and its value is determined purely by radiative corrections.


\section{Radiative corrections and vacuum alignment} 
\label{sec:quantumCorrections}


\noindent
The Yukawa terms and the gauging of the EW
symmetry break the global SU(4) symmetry explicitly.
This explicit breaking is reflected to the
scalar potential via quantum corrections, and the
true vacuum structure has to be determined by
taking loop corrections into account. This will determine
the preferred value of $ \phi $ and, hence, to
what degree EW symmetry is broken. In this section
we will show that the model can accommodate
a pGB-like Higgs boson along with a radiatively
induced Fermi scale originating from near
the SUSY-breaking scale.

	\subsection{The one-loop effective potential}
	
	\noindent
	The effective scalar potential at one-loop level can be written as
	\begin{equation}
	    \label{eq:Veff}
	    V_{\mathrm{eff}} = V_{\mathrm{tree}} + V_{\mathrm{1}}\,,
	\end{equation}
	where the tree-level potential, $ V_{\mathrm{tree}} = V_d + V_f + V_{\mathrm{soft}} $, is given by Eqs.~\eqref{eq:Vd2}, 
	\eqref{eq:Vf2}, and~\eqref{eq:softBreaking} respectively. In the $\overline{\mathrm{DR}}$ scheme the one-loop Coleman--Weinberg potential is given by\footnote{Beyond tree level, we have to include the Yukawa term contributions to the $ f $ term, as they influence the corrections even with vanishing vevs for the stop fields.}
	\begin{equation}
	    \label{eq:V1loop}
	    V_{\mathrm{1}} = \frac{1}{64\pi^2} \mathrm{Str}\left[\mathcal{M}^4(\Phi) \left(\log\frac{\mathcal{M}^2(\Phi)}
		{\mu_0^2}-\frac{3}{2}\right)\right],
	\end{equation}
	where ${\cal M}(\Phi)$ is the background-dependent tree-level mass matrix, 
	and the supertrace, $\mathrm{Str}$, is defined by
		\begin{equation}
		\mathrm{Str} = \sum_{\text{scalars}}-2\sum_{\text{Weyl fermions}}+3\sum_{\text{vectors}}. 
		\end{equation}	
	We fix the renormalization scale, $ \mu_0 $, so that $ \partial V_1/\partial \sigma =0 $. This usually gives a renormalization scale somewhere in between the values of the $ \sigma $ and $ \sing $ vevs, and keeps the effect of the radiative corrections small\footnote{There is nothing forcing this particular choice; in principle the $ \mu_0 $ can be chosen freely. Our choice of introducing a renormalization condition is mostly a matter of convenience as the vacuum conditions of Eq. \eqref{eq:eff_minimum_condition} are equivalent to enforcing $ \partial V_0 /\partial \sigma =0 $ at the full 1-loop vacuum. }.  The one-loop effective potential will then determine $ \brackets{\sing} = \tfrac{1}{\sqrt{2}}u $, $ \brackets{\sigma} = \tfrac{1}{\sqrt{2}}v $, and the EW embedding angle $ \phi $. These three quantities are found by requiring that the potential satisfies 
		\begin{equation}
		\dfrac{\partial V_{\eff}}{\partial \sigma} = 
		\dfrac{\partial V_{\eff}}{\partial \sing} = 0, \quad \text{and}\quad \dfrac{\partial V_{\eff}}{\partial \phi} = 0
		\label{eq:eff_minimum_condition}
		\end{equation} 
	at the vacuum. From Eq. \eqref{eq:Higgses} it follows that the Higgs vacuum of the model is $ \brackets{h_\uu^0}=\brackets{h_\dd^0}=-\tfrac{1}{2}v\sin\phi $. Therefore, the Fermi scale lies well below the actual symmetry breaking scale, $v$, if the embedding angle acquires a non-vanishing value $ 0<\phi \ll 1 $. 
	
    \subsection{Components of the neutral Higgs bosons}

	\noindent
	There are four real scalars contributing to 
	the CP-even part of the neutral Higgs components as given in Eq. \eqref{eq:Higgses}. These will in general mix in a non-trivial way, but in the limit $\phi\ll1$, the mass eigenstates approximately take the form\footnote{The complex scalars of the model are split into real fields like $ \eta = \tfrac{1}{\sqrt{2}}\left(\eta^R + \ii \eta^{I}\right) $ for any complex scalar $ \eta $.}
		\begin{eqnarray}
		\begin{pmatrix}	\varphi^R_1 \\ \varphi^R_2	\end{pmatrix} \simeq \begin{pmatrix} \cos \xi & \sin \xi \\ -\sin \xi & \cos \xi \end{pmatrix} \begin{pmatrix} \sing^R \\ \sigma^R	\end{pmatrix}, \nonumber \\ 
		h^0 \simeq \pi^{I}_4, \qquad \text{and} \qquad H^0 \simeq \pi_3^R.
		\label{eq:HiggsEigenstates}
		\end{eqnarray}	
	This identification holds exactly at tree level where three of the scalars are massive. The inclusion of corrections to the mass matrix is not expected to produce off-diagonal elements larger than $ m_{h^0}^2 $, and therefore the mass eigenstates are not changed significantly\footnote{With the parameters used in the next section, the deviation from this approximation is of order one in a thousand at the 1-loop level.}.    
	The lightest mass eigenstate, which is to be identified with the observed 125-GeV Higgs, $h^0$, consists almost exclusively of the pGB $ \pi^{I}_4 $. From Eq.~\eqref{eq:Higgses} we see that the couplings between $ h^0 $ and other SM particles are suppressed by only a factor of $ \cos \phi $ compared to the SM Higgs. For $\phi\ll 1$ this is well within current experimental bounds \cite{ATLAS:2015bea,Khachatryan:2014jba}. On the other hand, the couplings between $ \varphi^R_{1,2} $ and the SM particles are highly suppressed with a factor of $ \sin\phi \sin\xi $ and $ \sin\phi \cos \xi $, respectively, compared to the corresponding SM-Higgs coupling. 
	
	For comparison to the usual two-Higgs-doublet model (2HDM) of type II, we note that in the approximation used in Eq. \eqref{eq:HiggsEigenstates}, the CP-even Higgs bosons can be decomposed as
		\begin{align}
		\mathrm{Re} &\begin{pmatrix}	h_\uu^0 \\ h_\dd^0	\end{pmatrix} + \dfrac{\sin \phi}{2} \begin{pmatrix} v \\ v \end{pmatrix} \nonumber\\ &=\dfrac{1}{\sqrt{2}} \begin{pmatrix}\cos \alpha & -\sin \alpha \\ \sin \alpha & \cos \alpha 	\end{pmatrix} \begin{pmatrix} H^0 \\ h^0 \end{pmatrix} + \mathcal{O}(\phi),
		\end{align}  
	with $ \alpha = -\tfrac{\pi}{4} $ and $\tan\beta =1$. We find that the model naturally corresponds to the decoupling limit of 2HDM, with the only deviations being due to order-$\phi $ contributions to the neutral Higgs components coming from the additional $ \varphi_1^R $ and $ \varphi_2^R $ states \cite{Gunion:2002zf}. This ensures an SM-like low-energy behaviour of the model in spite of it containing two Higgs doublets.  
	
With the identification of the lightest Higgs from Eq. \eqref{eq:HiggsEigenstates}, the running Higgs mass can be determined from the effective potential approximately by
		\begin{equation}
		m_{h^0}^2 \simeq \left.\dfrac{\partial^2 V_{\mathrm{eff}}}{(\partial \pi_4^{I})^2}\right|_{\mathrm{vac}}
		\label{eq:approx_Higgs_mass}
		\end{equation}
	evaluated at the full vev. 
In the MSSM the top sector is the primary source for corrections to the Higgs mass, and this sector certainly contributes significantly here too, due to the strong coupling strength $ y_t $. However, in our model the other new particles of the Higgs sector are heavy, viz. in the multi-TeV range; in the numerical example presented later, many of the $M$-sector scalars are  heavier than the top squarks. For this reason especially the Higgs scalars play a  relevant role in the corrections to the lightest Higgs mass. 	
	
    \subsection{Spectrum}

	\begin{table}
		\centering
		\begin{tabular}{|lll|}
			\hline $ a = 0.860, $  & $ \lambda = 0.313, $  & $ m =\SI{3.50}{TeV}, $ \\ 
			$ m_{\LL\RR} = \SI{3.00}{TeV}, $  & $ m_{BW}=\SI{3.00}{TeV}, $ & $ \beta_{\mathrm{s}} = (\SI{2.75}{TeV})^2. $ \\ \hline\hline
			$ u = \SI{1.63}{TeV}, $ & $ v = \SI{5.63}{TeV} , $ & $ \phi = \num{0.0432}. $ \\ \hline
		\end{tabular}
		\caption{Using the parameters of the upper part of the table for our model, we find $ v = \sqrt{2}\brackets{\sigma} $, $ u= \sqrt{2}\brackets{\sing}$, and the EW-embedding angle for the resulting potential including one-loop corrections at renormalization scale $ \mu_0 = \SI{2.40}{TeV} $. Here we have taken all the dimensionful coupling constants $ b $, $ \mu $, and $ c $ in the superpotential to vanish.} 
		\label{tab:parameters}
	\end{table} 
	
	\noindent
	We will now provide the spectrum of the model via
a numerical analysis. The main challenge is determining
the minimum of the effective potential,
and we have performed that numerically. From
the (s)fermion contributions, we include the dominant
top corrections as the other (s)fermion contributions
are suppressed by the smallness of the
Yukawa couplings. This introduces two new parameters
to the one-loop analysis: $ m_\LL $ and $ m_\RR $,
the SUSY-breaking masses for the left- and right-handed
stops respectively. Furthermore, we include
the winos and the bino in the analysis with
the explicit SUSY-breaking masses $ m_W $ and $ m_B $.
The gluino does not contribute to the Higgs mass
nor the vacuum alignment at the one-loop level,
and its mass is, in principle, a free parameter.
However, assuming a single SUSY-breaking scale,
we expect the gluino mass to be around the wino
and bino masses.

	We expect that as long as the renormalization scale is chosen sensibly on the same order of magnitude as the vevs, the one-loop vevs should not change considerably from the tree-level values. Otherwise this will indicate that the perturbative procedure is unreliable. With this in mind, our search for an appropriate set of parameters starts with choosing a vev $ (u,v) $, which then determines $ \sin\phi = v_{\mathrm{EW}}/v $ in order for the model to reproduce the Fermi scale of the SM.  
	Furthermore the renormalization scale is chosen so that at the vacuum the $\sigma$-tadpole contribution vanishes, i.e.
		\begin{equation}
		\left.\dfrac{\partial V_1}{\partial\sigma}\right|_{\mathrm{vac}} = 0.
		\label{eq:renormalization_condtion}
		\end{equation}  
	Eqs. \eqref{eq:renormalization_condtion} and \eqref{eq:eff_minimum_condition} give four conditions that should be satisfied at the minimum allowing us to determine $ \mu_0 $, $ \beta_{\mathrm{s}} $, $ m $, and $ a $ dynamically for any choice of the other parameters $ \lambda $, $ m_{\LL\RR} $, $ m_{WB} $, and vacuum location. 

	The aim of our analysis is to determine whether there are points in the parameter space which can reproduce current experimental data, rather than doing an extensive analysis of the whole parameter space.  
	A set of benchmark parameters 
	is presented in Table~\ref{tab:parameters}. 
	In the numerical analysis, we have taken into account the one-loop running of the SM parameters to the renormalization scale.
	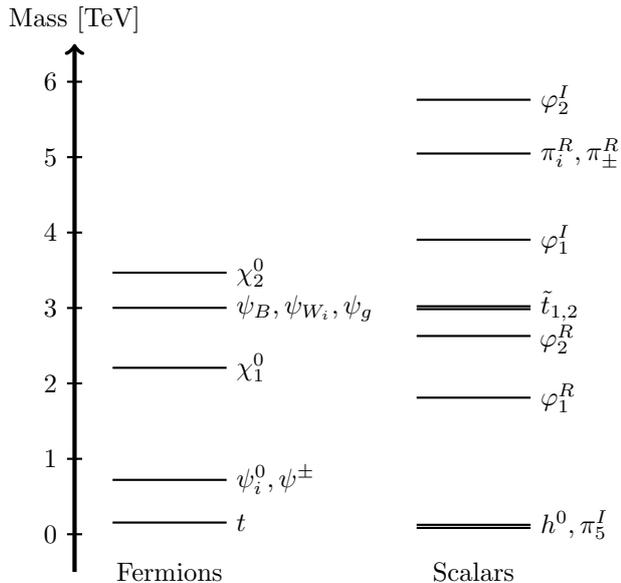
\begin{figure}
		\centering
		\begin{tikzpicture}
		\node[label] at (4.75,-0.5) {Scalars};
		
		\draw[thick] (4,0.084) -- (5.5,0.084); 
		\draw[thick] (4,0.125) -- (5.5,0.125) node[right] {$ h^0, \pi^{I}_5 $};
		\draw[thick] (4,1.811) -- (5.5,1.811) node[right] {$ \varphi_1^{R} $};
		\draw[thick] (4,2.629) -- (5.5,2.629);
			\node[label] at (5.85,2.591) {$ \varphi_2^{R} $}; 
		\draw[thick] (4,2.985) -- (5.5,2.985);
		\draw[thick] (4,3.023) -- (5.5,3.023) node[right] {$ \tilde{t}_{1,2} $};		
		\draw[thick] (4,3.905) -- (5.5,3.905) node[right] {$ \varphi^{I}_1 $};
		\draw[thick] (4,5.048) -- (5.5,5.048) node[right] {$ \pi_i^{R}, \pi_{\pm}^{R} $};
		\draw[thick] (4,5.761) -- (5.5,5.761) node[right] {$ \varphi_2^{I} $};
		
		\node[label] at (0.75,-0.5) {Fermions};
		
		\draw[thick] (0,0.155) -- (1.5,0.155) node[right] {$ t $};
		\draw[thick] (0,0.721) -- (1.5,0.721) node[right] {$ \psi_i^0, \psi^{\pm} $};
		\draw[thick] (0,2.207) -- (1.5,2.207) node[right] {$ \chi_1^0 $};
		\draw[thick] (0,3.002) -- (1.5,3.002) node[right] {$\psi_B, \psi_{W_i}, \psi_g $};
		\draw[thick] (0,3.468) -- (1.5,3.468) node[right] {$ \chi_2^0 $};
		
		\draw[ultra thick, ->] (-0.5,-0.5) -- (-0.5,6.5) node[above] {Mass [TeV]} ;
		\foreach \y/\j in {0/0,1/1,2/2,3/3,4/4,5/5,6/6} \draw[thick] (-0.4,\y) -- (-0.6,\y) node[left] {\j};
		\end{tikzpicture} 
		\caption{A schematic representation of the (running) mass spectrum contained in the model resulting from the parameters of Table \ref{tab:parameters}. {\bf Fermions}: $ t $ is the top quark; $ \psi_i^{0} $ and $ \psi^{\pm} $ are the fermionic partners of the GBs; $ \chi^0_{1,2} $ are the mass eigenstates of the fermionic partners of $ \sigma $ and $ \omega $; $ \psi_B $, $ \psi_{W_i} $, and $ \psi_g $ are the binos, winos, and gluinos respectively. {\bf Scalars}: $ \pi_5^{I} $ is the singlet pGB; $ h^0 $ is the 125-GeV Higgs boson; $ \varphi_{1,2}^{R,I} $ are the mass eigenstates of the $ \sigma  $ and $ \omega $; $ \tilde{t}_{1,2} $ are the top squarks; $ \pi^R_i, \pi_{\pm}^R $ are the real parts of the five $ \pi $ scalars.       
		}
		\label{fig:spectrum}
	\end{figure}
	
	Using the parameters presented in Table \ref{tab:parameters} our model is able to reproduce the Fermi scale of the SM with $ v_{\mathrm{EW}} = v \sin \phi = \SI{246}{GeV} $, and a lightest Higgs scalar with a running mass of $ \SI{125}{GeV} $. The spectrum of the model with the current set of parameters is presented schematically in Figure \ref{fig:spectrum}. This is a tree-level spectrum except for the lightest Higgs boson, whose mass stems exclusively from radiative contributions. The lightest new scalar is $ \varphi_1^R $ at \SI{1.8}{TeV}, while new fermions, $ \psi_i^0 $ and $ \psi^{\pm} $, are found already at \SI{700}{GeV}. These roughly correspond to the Higgsinos of the MSSM and obtain their masses from the  singlet vev in the absense of the $\mu$ term. The mixing angle from Eq. \eqref{eq:HiggsEigenstates} will in this case take the value $ \xi = 0.41 $, indicating some mixing between $ \sigma^R $ and $ \sing^R $, and a $\varphi_1^R$ that is dominated by $ \sing^R $. Of the remaining new particles of the $M$ sector, we see that most are spread out with masses between 2 and \SI{6}{TeV} with the exception of $ \pi^{I}_{1,2,3} $ and $ \pi^{I}_5 $. The $ \pi^{I}_{1,2,3} $ are the GBs of the EWSB and are absorbed into the gauge bosons. The last scalar, $ \pi^{I}_5 $, is a pGB getting a radiatively generated running mass of about \SI{80}{GeV} for the parameters used here. At the tree level $ \pi_5^{I} $ couples directly only to other $ M$-sector particles. Furthermore $ \pi_5^{I} $ is protected from decay by a $ Z_2 $ symmetry, so it will only show up in colliders as at least \SI{160}{GeV} of missing energy. Due to the $Z_2$ symmetry it is also an interesting candidate for dark matter (DM), but the study of DM phenomenology is left for future work.

	\subsection{Mass of the lightest Higgs boson}
	
	\noindent
	Our analysis is constrained to yield a running Higgs mass of $ m_{h^0} = \SI{125}{GeV} $. Although it is involved to determine the contributions to its mass, it arises radiatively because of the breaking of the global $\SU(4)$ symmetry due to the couplings between the $ M $-sector and the SM fields. 
	In the MSSM and NMSSM  the dominant radiative contributions are typically due to (s)top particles due to the strong Yukawa coupling~\cite{Martin:1997ns,Ellwanger:2009dp}. In the NMSSM, however, at small $\tan\beta$, there is a large tree-level contribution to the Higgs mass due to the additional singlet. This increases the upper limit for the Higgs mass compared to the MSSM substantially~\cite{Ellwanger:2006rm,Bagnaschi:2014rsa}. We expect a similar effect in our case due to the departure from the symmetry limit induced by the radiative effects.	

In addition to the (s)tops, all the new heavy particles shown in Fig.~\ref{fig:spectrum} give appreciable contributions to the Higgs mass due to the $\SU(4)$-breaking sectors. This can be understood in terms of Eqs. \eqref{eq:V1loop} and \eqref{eq:approx_Higgs_mass}, giving 
		\begin{align}
		m^{2}_{h^0} \simeq &\dfrac{1}{64\pi^2}\mathrm{Str}\left[2 \left(\dfrac{\partial \mathcal{M}^2}{\partial\pi_4^{I}}\right)^2\right. \nonumber \\
		&\quad +\dfrac{\partial^2 \mathcal{M}^4}{(\partial \pi_4^{I})^2}\left(\log \dfrac{\mathcal{M}^2}{\mu_0^2}-1\right) \Bigg]_{\mathrm{vac}}.
		\end{align}
	In the expression above we retained only the relevant terms. 
	It shows that the contribution coming from each mass eigenstate is roughly proportional to the square of the coupling between the corresponding particle and the Higgs boson  times the square of the mass of this particle. If other particles are heavier than the (s)top, this can be enough to compensate for a weaker coupling. This explains  why we have other corrections comparable to  the ones due to the (s)top sector. 
	 
	We note that the model parameters require some amount of tuning to achieve the right value for the Fermi scale and Higgs mass through the radiative corrections. This tuning is similar to what one would find in the MSSM at a comparable SUSY-breaking scale. This is due to the (s)top corrections to the Higgs mass being functionally identical in the two models; only the coupling strength of the the Higgs boson to the (s)top is slightly modified. 

\begin{figure}
    \begin{center}
	\includegraphics[width=0.45\textwidth]{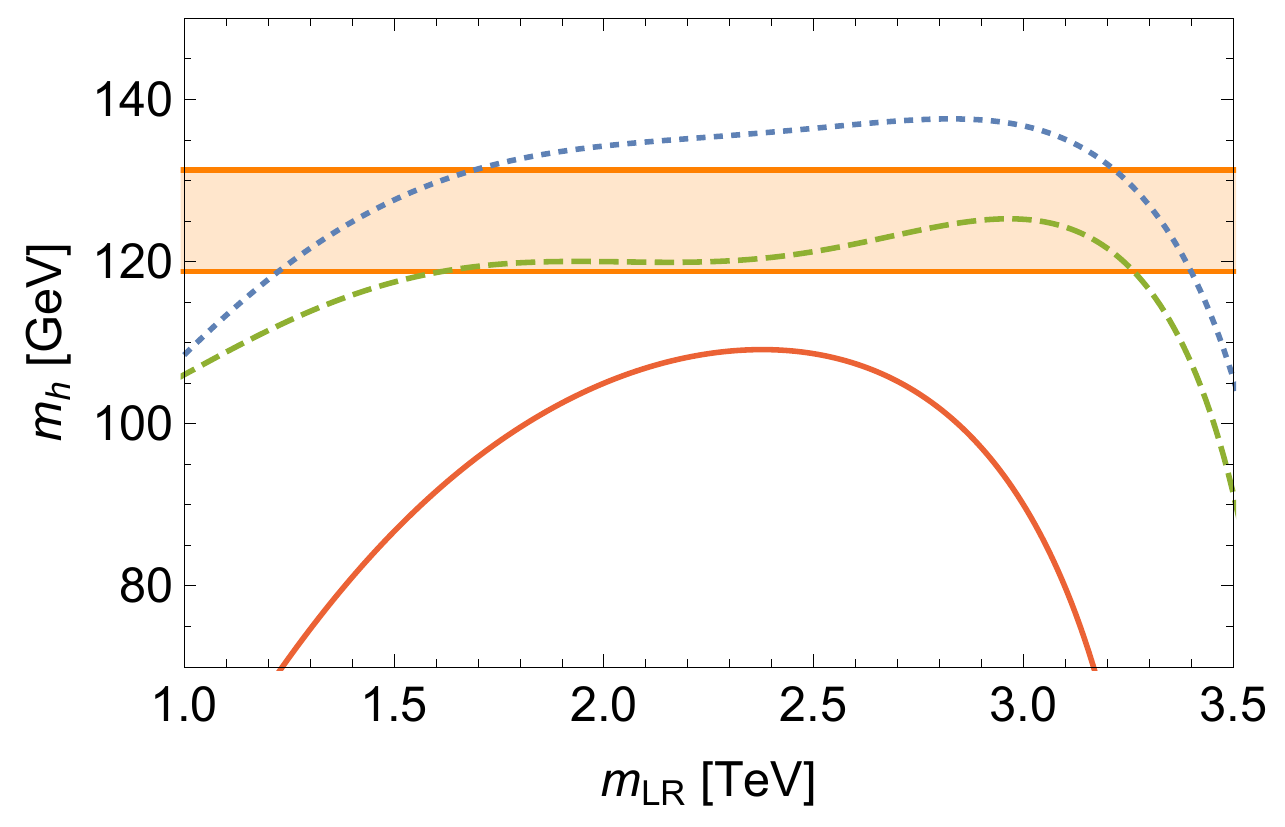}
    \end{center}
    \caption{The Higgs mass as a function of the soft-breaking stop mass parameter, $m_{\mathrm{LR}}$, for three different sets of model parameters\\
    $(a, \lambda, m_{BW}, \beta_{\mathrm{s}})=(0.86, 0.34, 3.0\, \mathrm{TeV},9.5\, \mathrm{TeV}^2)$,\\
    ~\hspace*{2.8cm}$(0.86, 0.313, 3.0\, \mathrm{TeV}, 7.6\, \mathrm{TeV}^2)$,\\
    ~\hspace*{2.8cm}$(0.90, 0.38, 3.0\, \mathrm{TeV}, 8.1\, \mathrm{TeV}^2)$, \\
    represented by solid, dashed and dotted curves respectively. Here we use a fixed renormalisation scale $\mu_0=2.4$ TeV. Simultaneously to the changing $m_{\LL\RR}$, we adjust the soft-breaking parameter $m$ such that the EW spectrum remains fixed. The shaded band represents 5\% deviation from the Higgs mass 125\,GeV to acknowledge the difference between the physical and $\overline{\mathrm{DR}}$ mass, and a theoretical uncertainty due to the missing higher-order corrections. }
    \label{fig:hmasses}
\end{figure}

To illustrate that the parameter space can accommodate a relatively wide range of Higgs-mass values, in Fig.~\ref{fig:hmasses} we show the Higgs mass as a function of the soft-breaking stop mass parameter, $m_{\mathrm{LR}}$, for three representative sets of model parameters. We simultaneously adjust the soft-breaking parameter $m$ such that the EW spectrum remains fixed. For this illustration we use a fixed renormalisation scale 2.37 TeV.
The dashed curve in the figure coincides with our benchmark parameter set for the stop mass parameter of 3 TeV. 

	For a more precise determination of the mass of $ h^0 $ in our model, corrections beyond one-loop level can be relevant.  For example in  (N)MSSM  fixed-order related studies, one observes that for stop masses at around 1 TeV, $\mathcal O(m_t^2 y_t^2 \alpha_s)$ corrections lead to mass changes of about 10 to 20 GeV  depending on the parameter point and the chosen renormalization scheme and scale~\cite{Heinemeyer:1998np,Espinosa:1999zm, Degrassi:2001yf,Heinemeyer:2007aq,Muhlleitner:2014vsa}. Further corrections of $\mathcal O(m_t^2 y_t^4)$ can also induce a mass shift of several GeV \cite{Espinosa:2000df,Brignole:2001jy}. Furthermore if the top-squark masses are large, then a pure fixed-order calculation  must be amended by resumming large logarithms. These higher-order analyses go beyond the present explorative scope of our work but do not modify the overall scenario: since the numerical analysis has shown that the model can accommodate a broad range of different Higgs boson masses, cf. Fig.~\ref{fig:hmasses}, higher-order corrections should not change the viability of the model.

\section{Conclusions}
 \label{sec:conc}
\noindent
By combining the elementary-Goldstone-Higgs scenario with SUSY, we have demonstrated, via an explicit determination of the quantum ground state of the model, that it is possible to raise the scale of supersymmetry breaking to several TeVs while retaining the relevant qualities of a supersymmetric extension of the SM. 

We achieved this by first enhancing the global symmetries of the super-Higgs sector to SU(4) and then embedding the electroweak sector including the chiral superfields of the SM fermions. We then investigated the minimal requirements needed to achieve a vacuum of the model such that the resulting Higgs is a pGB. The spontaneous-symmetry-breaking scale of $\SU(4)$, also identified with the SUSY-breaking scale, is several TeVs above the radiatively induced Fermi scale. Because of the enhanced SU(4) symmetry, the model  predicts the existence of two super-Higgs multiplets with one mass eigenstate playing the role of the pGB.
The symmetry-breaking structure also implies $\tan \beta =1$ and the need of a supplementary singlet chiral superfield. 

The spectrum of the superpartners of the SM fermions and gauge bosons lies in the multi-TeV range, thereby complying with the lack of observation of (sub-)TeV states expected in Fermi-scale SUSY-breaking scenarios.
Because the particle content of the model essentially resembles the (N)MSSM particle content from the grand-unification perspective, we expect analogous unification scenarios to be viable in this framework as well, albeit with modified threshold corrections.

\acknowledgments

\noindent
The $\mathrm{CP}^3$-Origins centre is partially funded by the Danish National Research Foundation, grant number DNRF90. HR acknowledges support by the National Science Foundation under Grant No. NSFPHY11-25915. TA acknowledges partial funding from a Villum foundation grant. 

\clearpage
 
\appendix

\bibliography{susy.bib}

\end{document}